# Multilayer Silicene: a clear evidence

Paola De Padova*[1, 2], Amanda Generosi[2], Barbara Paci[2], Carlo Ottaviani[2], Claudio Quaresima[2], Bruno Olivieri[3], Eric Salomon[4], Thierry Angot[4] and Guy Le Lay[4]

[1] Aix Marseille Univ, IMERA, Marseille, France

[2] Consiglio Nazionale delle Ricerche-ISM, via Fosso del Cavaliere 100, 00133 Roma, Italy

[3] Consiglio Nazionale delle Ricerche-ISAC, via Fosso del Cavaliere 100, 00133 Roma, Italy

[4] Aix Marseille Univ, CNRS, PIIM, UMR 7345, Marseille, France

*Corresponding author: depadova@ism.cnr.it

Abstract

One year after the publication of the seminal paper on *monolayer* 3×3 reconstructed silicene grown on a silver (111) substrate, evidence of the synthesis of epitaxial √3×√3 reconstructed *multilayer* silicene hosting Dirac fermions was presented. Although a general consensus was immediately reached in the former case, in the latter one, the mere existence of *multilayer* silicene was questioned and strongly debated. Here, we demonstrate by means of a comprehensive x-ray crystallographic study, that multilayer silicene is effectively realized upon growth at rather low growth temperatures (~200°C), while, instead, three-dimensional growth of silicon crystallites takes place at higher temperatures, (~300°C). This transition to bulk like silicon perfectly explains the various data presented and discussed in the literature and solves their conflicting interpretations.

Despite a visionary theoretical paper originating in 1994 [1], ten years before the advent of graphene, followed by a couple of others more than twelve years later [2,3], the real possibility of synthesizing silicene, the silicon based analogue of graphene, was considered with great scepticism. This largely shared opinion has radically changed in 2012, after the first compelling evidence of the realization of *monolayer* silicene, in the seminal, 3×3 reconstructed, phase on a silver (111) substrate, coinciding with a 4×4 Ag(111) supercell [4], quickly followed by the new evidence of the formation of √3×√3 reconstructed monolayer silicene on a zirconium diboride thin film [5]. Since then, "The growth and properties of silicene" has become one the 10 "hottest research fronts" in physics according to a citation-based study of 2014 by Thomson- Reuters, being topic number one in condensed matter physics [6]. Indeed, the realization of *monolayer* silicene has widened the horizon with a cornucopia of new exotic properties hardly accessible for graphene, like, e.g., the quantum spin Hall effect [7-9]. It has further paved the way to the synthesis of novel artificial two-dimensional (2D) elemental materials, namely, germanene in 2014 [10], stanene [11] and borophene [12], in 2015.

Contradicting a pessimistic prediction on the practical usability of silicene [13], field-effect transistors operating at room temperature made with a *monolayer* silicene channel were fabricated in 2015 [14].

M*ultilayer* silicene was theoretically envisaged in 2013 [15,16]. The first evidence of its realization on top of the initial archetype 3×3 *monolayer* silicene on Ag(111) appeared the same year [17]. The growth proceeds in successive flat terraces separated by ~0.3 nm, which all show a unique √3×√3 reconstruction [18]. The size of the √3×√3 cell, as measured in scanning tunneling microscopy (STM) is ~0.64 nm, which points to a ~3.8 % contraction with respect to the initial unit cell of the 3×3 reconstructed *monolayer* silicene on Ag(111) [18,19]. Key new features associated with *multilayer* silicene are, for one, the emergence of Dirac fermions, clearly manifested in angle-resolved photoelectron spectroscopy (ARPES) measurements [17] and in scanning tunneling spectroscopy (STS) ones [19], and, for two, the stability of the multilayer film in ambient air, protected by its ultrathin native oxide, for at least 24 hours [20]. The unique fingerprint of *multilayer* silicene was unravelled in an *ex-situ* Raman spectroscopy study, which revealed a ~ (3.2 ± 0.3) cm$^{-1}$ blue shift of the main line with respect to that of a piece of Si(111) wafer [20].

Unfortunately, at the present time, no clear theoretical support for such *multilayer* silicene film has been gained. An interesting tentative has been the proposal of dumbbell moieties, building so-coined silicite, a layered allotrope of silicon [21], but its calculated structure, although giving the correct lateral contraction, gives a much too large vertical separation between successive layers.

On the experimental side, a strong debate has developed between supporters of *multilayer* silicene and their opponents. Typically, the pros base their favourable conclusion on STM observations, and, further, on *in situ* electrical measurements [18] or *in situ* Raman studies [22]. Instead, the cons, argue that the √3×√3 structure stems from silicon (111) islands covered by a monolayer of Ag atoms arranged, at room temperature (RT), in the well-known honeycomb chained triangle (HCT) Si(111)√3×√3-Ag reconstruction [23,24]. They base also their argumentation on STM observations [25] and Raman spectroscopy [26], and, further, on LEED intensity studies [27,28], LEEM [29] and TEM observations [26], ARPES studies [30], as well as Auger electron spectroscopy (AES) and optical measurements [31]. However, in most cases these experiments have been carried out at temperatures significantly higher than 200°C, meaning that, at variance with the works of the pros, the initial layer is not the sole archetype 3×3 phase, but, often, instead, a patchwork of different phases, locally more or less ordered [32].

A third intermediate possibility has been recently documented since the formation of a Si(111)(√3×√3)R30°-Ag reconstruction on the surface was distinctively ruled out by peeling off the surface layer with the STM tip [33]. *Multilayer* silicene would not exist *per se*, but epitaxial bulk-like Si(111) films with spontaneous intrinsic √3×√3 honeycomb superstructure would form, showing delocalized surface state as well as linear energy-momentum dispersion observed from quasiparticle interference patterns [33]. In this respect we recall that years ago, Fan *et al.*, had found an unusual intrinsic Si(111)√3×√3 reconstruction, which they interpreted as a vacancy model, i.e., as a matter of fact, consisting in a surface honeycomb structure [34, 35]. Amazingly, upon direct comparison, the measured LEED intensity I(V) spectra were strikingly similar to those of the Si(111)√3×√3-Ag surface.

To address these highly controversial issues, we have undertaken a comprehensive study of the system, comparing the Si(111)√3×√3-Ag surface used as a reference, with Si growth on Ag(111) crystals, either in the low temperature regime (at ~200°C), or in the high temperature one (~300°C). To unveil the nature of the resulting nanomaterials, we have carried out, for the first time, an accurate ED-GIXRD study. Along with STM, LEED, AES and Raman characterizations, it has permitted to gain a clear picture. At ~200°C, *multilayer* silicene with no Ag surface layer on top and a layer cell size of a = (6.477 ± 0.015) Å, totally different from that of Si(111), $a_{Si}$ = (3.842 ± 0.003) Å, develops. Instead, at ~300°C, the surface lattice parameter is measured at (6.655 ± 0.015) Å, like in the case of the Si(111) surface terminated by the Si(111)√3×√3-Ag reconstruction, i.e., with one monolayer (1 ML) of silver on top, which indicates the formation of Si(111) crystallites.

The Ag(111) substrate surfaces were prepared by $Ar^+$-bombardment (2 kV, $5 \times 10^{-5}$ mbar) and subsequent annealing at ~550 °C for about 30 minutes of (111)-oriented Ag single crystals under ultra-high vacuum conditions (base pressure $0.9 \times 10^{-10}$ mbar). Sharp 1×1 low-energy electron diffraction (LEED) patterns were observed and Auger electron spectroscopy measurements using a double pass cylindrical mirror revealed no trace of carbon, oxygen or other contaminants. Silicon was deposited at a ~ 0.03 ML/min rate from a source consisting of a directly heated Si-wafer piece. The reference Si(111)√3×√3-Ag sample was obtained by depositing one monolayer (ML) of silver from a W crucible at a rate of 0.05 ML/min, onto a clean Si(111)7×7 crystal at ~500 °C [24].

STM images were acquired by using a commercial Omicron VT-STM. Homemade STM tips were fabricated from electrochemically etched tungsten wires in 2 M NaOH solution. STM images were recorded at room temperature, in constant current mode and processed using the WS×M software [36]. Linear electronic drift correction was systematically used to compensate for possible thermal and mechanical drift of the probe.

ED-GIXRD measurements were performed in room conditions in reflection mode $\theta_i = \theta_r$ using a Bremsstrahlung X-ray radiation ranging up to 50 keV produced by a W-anode X-ray tube on a non-commercial ED spectrometer [37-40]. The schematic layout of the ED-GIXRD set-up is reported in Fig. SM1 in the Supplemental Materials. For all studied samples, in order to minimize the diffracted signal arising from the monocrystalline substrates, a preliminary rocking curve procedure was performed (in plane tilt) to set the optimal $\alpha$ value and subsequently a combined out of plane $\psi$ tilt and a $\phi$ rotation was performed to collect the in plane structural information. This goal was achieved using a custom designed three axes sample holder used to perform the $\alpha$, $\psi$, $\phi$ tilts and rotation.

Fig. 1a) displays LEED patterns and AES spectra for 1ML of silicon (bottom) deposited onto the Ag(111) surface at ~200°C, which forms the archetype 3×3 reconstructed silicene layer, exactly matching a silver 4×4 supercell and for 10 MLs deposited at the same temperature (top), which show an apparent √3×√3 reconstruction with respect to monolayer silicene. In the former case, the ratio between the Ag and Si AES signals, $I_{Ag}/I_{Si}$ is 1.16, while it is 0.09 in the latter. This reveals imperfect wetting of the formed 10 ML film, in accord with the growth in successive flat terraces [18], as illustrated in the STM image of Fig. 1b), where a line profile is traced on the √3×√3 film (red line). When Si deposition is performed at ~300°C the ratio is 0.53, as shown in Fig. SM2, instead of 0.09 when performed at ~200°C, which points to strong silicon clustering.

At first diffraction measurements (E = 50 keV) were performed on the reference Si(111)√3×√3-Ag sample, prepared in situ (see Fig. SM2), then collecting ED-GIXRD patterns at different scattering $\theta$ angles (2.15°; 2.65°; 3.00°; 3.80° and 4.60°) in order to maximize the explored q-region measuring the whole reciprocal space of interest. The results are shown in figure 2. The following crystalline reflections were detected: Si(111) $q_z = (2.003 \pm 0.005)$ Å$^{-1}$, $d_{111} = (3.136 \pm 0.007)$ Å Si(222) $q_z = (4.008 \pm 0.005)$ Å$^{-1}$, $d_{222} = (1.568 \pm 0.003)$ Å ; Si(220) $q_{x,y} = (3.270 \pm 0.005)$ Å$^{-1}$, $d_{220} = (1.921 \pm 0.003)$ Å,

directly yielding the in-plane Si(111) lattice parameter $a_{Si(111)} = 3.842$ Å. They are in perfect accord with the JCPDS N. 27-1402 and JCPDS N. 04-0783 powder diffraction database. Most importantly, the Si(111)√3×√3-Ag cell was directly observed in a) at $q_{xy} = (0.944 \pm 0.005)$ Å$^{-1}$, corresponding to (6.655 ± 0.015) Å, i.e., in perfect agreement to √3 times $a_{Si(111)}$. In b), the $q_z$ component is obviously dominant due to the geometry of this experiment (tilts: α = 0.050°, ϕ =0.050°). The internal in-plane structure of the reconstructed Si(111)√3×√3-Ag surface was also observed, showing its stability in air, at $q_{x,y} = (2.209 \pm 0.005)$ Å$^{-1}$, $d_{Ag-Ag} = (2.845 \pm 0.005)$ Å and $q_{x,y} = (2.925 \pm 0.005)$ Å$^{-1}$, $d_{Si-Si} = (2.148 \pm 0.006)$ Å in excellent agreement with the atomic distances between the Ag triangles and Si trimers constituting the anisotropic HCT reconstruction at RT [23, 41]. This confirms the accuracy of the collected data.

Next, the same experimental procedure was performed on the 10 MLs √3×√3 film prepared at ~200°C (see Fig. SM1) fixing the instrumental geometry at θi =θr = 2.50° and keeping E = 50 keV. Indeed, these conditions were found to be ideal to observe the desired q-region and minimize the fluorescence line disturbance, so that the K$\alpha_{1,2}$ and K$\beta_1$ Ag lines arising from the substrate did not overlap with any diffraction signal. In Fig. 3 a pattern is shown, as a representative of several measurements performed. Reflections were observed and attributed to the in plane x,y contributions of the multilayer film, namely, a first order in plane reflection at $q_{xy} = (0.970 \pm 0.005)$ Å$^{-1}$ corresponding to $a_{ML} = (6.477 \pm 0.015)$ Å and another at $q_{xy} = (1.939 \pm 0.005)$ Å$^{-1}$, corresponding to an in plane "second order" perfectly compatible with the terrace-like growth on Ag(111). In addition, an out of plane reflection is visible at $q_z = (2.033 \pm 0.005)$ Å$^{-1}$, corresponding to $d_{zML} = (3.090 \pm 0.010)$ Å, as previously reported for "multilayer silicene" [20]. Furthermore, the Ag(111) reflection at $(2.667 \pm 0.005)$ Å$^{-1}$, ($d_{zAg} = 2.356 \pm 0.005$ Å) and the in plane (200) reflection at $q_{xy} = (3.090 \pm 0.005)$ Å$^{-1}$ ($a_{Ag} = 4.067 \pm 0.002$ Å) were also collected, as expected. Note, that the relative intensities of the two peaks are not directly related to the crystallinity of the material since they are influenced by the α and ψ tilts corresponding to a quantitatively different projection of the momentum transfer on the xy plane or the z direction. Particularly noteworthy is that absolutely no Si(220) reflection was detected. This proves that the whole body of the film below the oxidized top layer possesses the $a_{ML}$ = (6.477 ± 0.015) Å in-plane lattice parameter, totally at variance with a bulk-like Si(111) arrangement terminated by a Si(111)√3×√3-Ag reconstruction.

We present in table I the direct comparison between these results.

| Si(111)√3×√3-Ag | 10 MLs √3×√3 film grown at ~200°C on Ag(111) | Si(111)√3×√3-Ag | 10 MLs √3×√3 film grown at ~200°C on Ag(111) |
|---|---|---|---|
| $q_{xy}$ | $q_{xy}$ | $q_z$ | $q_z$ |
| (0.944 ± 0.005) Å$^{-1}$ | (0.970 ± 0.005) Å$^{-1}$ | (2.003 ± 0.005) Å$^{-1}$ | (2.033 ± 0.005) Å$^{-1}$ |
| $\Delta q_{xy}/q_{xy}$ = 0.01 | $\Delta q_{xy}/q_{xy}$ = 0.01 | $\Delta q_z/q_z$ = 0.005 | $\Delta q_z/q_z$ = 0.005 |
| $a_{Si(111)\sqrt{3}}$ | $a_{ML}$ | $d_{zML}$ | $d_{Si(111)}$ |
| (6.655 ± 0.015) Å | (6.477 ± 0.015) Å | (3.136 ± 0.010) Å | (3.090 ± 0.010) Å |
| $\Delta d_{xy}/d_{xy}$=0.0045 | $\Delta d_{xy}/d_{xy}$ = 0.0045 | $\Delta d_z/d_z$=0.0064 | $\Delta d_z/d_z$=0.065 |

Table I: ED-GIXRD collected scattering vectors $q_{xy}$ (in plane) / $q_z$ (out of plane) and corresponding lattice parameters / vertical separations d. for the reference Si(111)√3×√3-Ag sample and for the multilayer film grown at ~200°C on Ag(111).

It is obvious that the √3×√3 multilayer film differs both from crystalline bulk Si(111) and from the Si(111)√3×√3-Ag reconstruction itself. We emphasize the significant 2.7% in-plane contraction with respect to Si(111)√3×√3-Ag, yet, with no vertical expansion (instead, a very small vertical contraction is noticed) in good agreement with the value of (6.44 ± 0.07) Å reported on the line profile in Fig. 1b, considering the surface lattice parameter of the 3×3 silicene phase to be 11.56 Å, and the literature [18,19]. Hence, these diffraction results rule out that the √3×√3 multilayer film grown on top of Ag(111) at ~200°C corresponds to the formation of a diamond-like crystalline Si film terminated by the Si(111)√3×√3-Ag reconstruction, with silver atoms acting as surfactant during the growth process [42].

The same experiments were reproduced, but on samples held at ~300°C during Si deposition onto Ag(111) with also 10 MLs, as shown in Fig. SM2. The ED-GIXRD pattern shown in Fig. 4 reveals the presence of crystalline Si(111) with $q_z = (2.003 \pm 0.005)$ Å$^{-1}$, $d_z = (3.137 \pm 0.005)$ Å; the in-plane (220) and (222) reflections were also detected. Moreover an in plane reflection at $q_{x,y} = (0.944 \pm 0.005)$ Å$^{-1}$, corresponding to a lattice parameter of $(6.655 \pm 0.015)$ Å, exactly that of the Si(111)√3×√3-Ag reconstruction, was measured. Hence, differently from the growth at low temperatures (~200°C), at high growth temperatures (~300°C), Ag acts as a surfactant in the formation of diamond-like crystalline Si terminated by the Si(111)√3×√3-Ag reconstruction.

All this is in nice agreement with the AES intensity ratios, $I_{Ag}/I_{Si} = 0.09$ and $I_{Ag}/I_{Si} = 0.53$, respectively at ~200°C and ~300°C. Moreover, the distinct nature of the nano materials formed is confirmed upon *ex-situ* Raman spectroscopy measurements, as shown in Fig. 5. Both the reference Si(111)√3×√3-Ag sample and the 10 ML sample grown at ~300°C show the characteristic bulk silicon line at 520.3 cm$^{-1}$, while, instead, the 10 ML sample grown at ~200°C presents a $(3.2 \pm 0.3)$ cm$^{-1}$ blue shift (Raman peak at 523.5 cm$^{-1}$) in agreement with ref. 20. This confirms our initial assignment of such films prepared at low temperatures (~200°C) to *multilayer* silicene, since it is a reasonable choice to give this name to the new Si-based 2D material, which we have first synthesized upon continuing growth on the first 3×3 silicene monolayer [17,18].

The strong debate between the pros and cons of the existence of *multilayer* silicene should be thus closed: *multilayer* silicene, indeed a metastable phase, is synthesized only in the low temperature regime, while in the high temperature one, naturally, Si(111) terminated by the Si(111)√3×√3-Ag reconstruction grows. Hence, when working in this high temperature regime, as most of the authors do, it is correct to state that just Si(111)√3×√3-Ag is formed in such conditions [26]. However, it is ambitious to conclude as a general statement, that multilayer silicene cannot be obtained [29, 30, 43].

Seemingly in contradiction, is the work of Shirai *et al.*, who worked in the low temperature regime [27]. Nevertheless, the authors did not use a silver single crystal as a substrate. Instead, they grew at RT a thin epitaxial Ag(111) film on a piece of Si(111) wafer and heated it to 230°C for Si deposition. But, they did not recall that such a procedure leads to the clumping of the smooth Ag film (displaying streaks in RHEED) into 3D Ag(111) islands (spotty RHEED pattern appearing as early as ~100°C [44]), leaving behind, beyond about 200°C, the Si(111)√3×√3-Ag surface reconstruction on the denuded zones [24].

An apparent exception, with experiments performed in the low temperature regime, is the work of Borensztein et al. [31]. Yet, we stress that for simplicity, they improperly adopted a laminar growth model, which is at variance with the well-established terrace growth [18, 20, 33, 45]. This clearly makes their interpretation that the surface reconstruction is induced by a surfactant layer of silver atoms during the growth of a silicon (111) thin film instead of being due to pristine multilayer silicene, highly questionable.

To summarize, we have conducted studies combining complementary tools indicating that *multilayer* silicene is successfully synthesized on silver (111) surfaces when growth is performed in the low temperature regime, typically at about 200°C. We have further confirmed that when growth is performed in the high temperature regime, instead, bulk-like Si(111) terminated by the Si(111)√3×√3-Ag reconstruction is formed. This clear-cut demonstration settles the issue, demonstrating that both structures can be observed and that they critically depend on the growth conditions.

**Acknowledgments**

Paola De Padova wishes to thank the IMERA (Aix-Marseille University) for the fellowship supporting her work from September 2015 to July 2016. We warmly thank Marco Guaragno, whose invaluable technical support permitted the feasibility of the ED-GIXRD measurements.

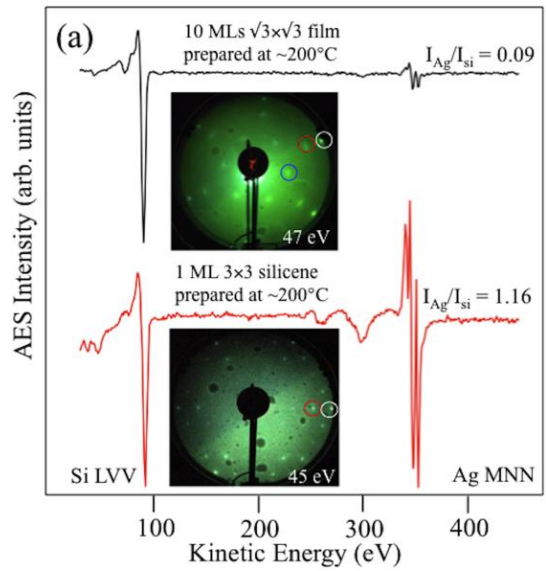

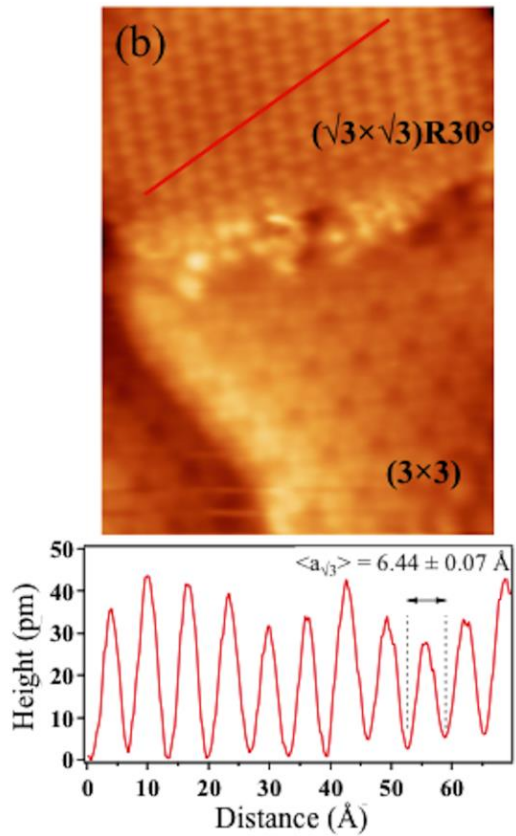

**Figure 1**. (a) Bottom: LEED pattern and AES spectrum for 1ML of silicon deposited onto the Ag(111) surface at ~200°C forming the 3x3 reconstructed silicene monolayer; white and red circles: silver and silicene integer order spots. Top: same for 10 MLs deposited also at ~200°C, which display a √3×√3

LEED pattern, blue circles: (1/3,1/3) order spots with respect to integer (1,1) silicene ones. The respective ratios between the Ag and Si AES signals, $I_{Ag}/I_{Si}$, are indicated. (b) Left: 90 Å × 125 Å filled-states STM image of the initial 3×3 silicene structure on Ag(111) Si (I = 140 pA, V = -49 mV) and of the √3×√3 film structure. Right: line profile along the red line traced on the √3 × √3 structures.

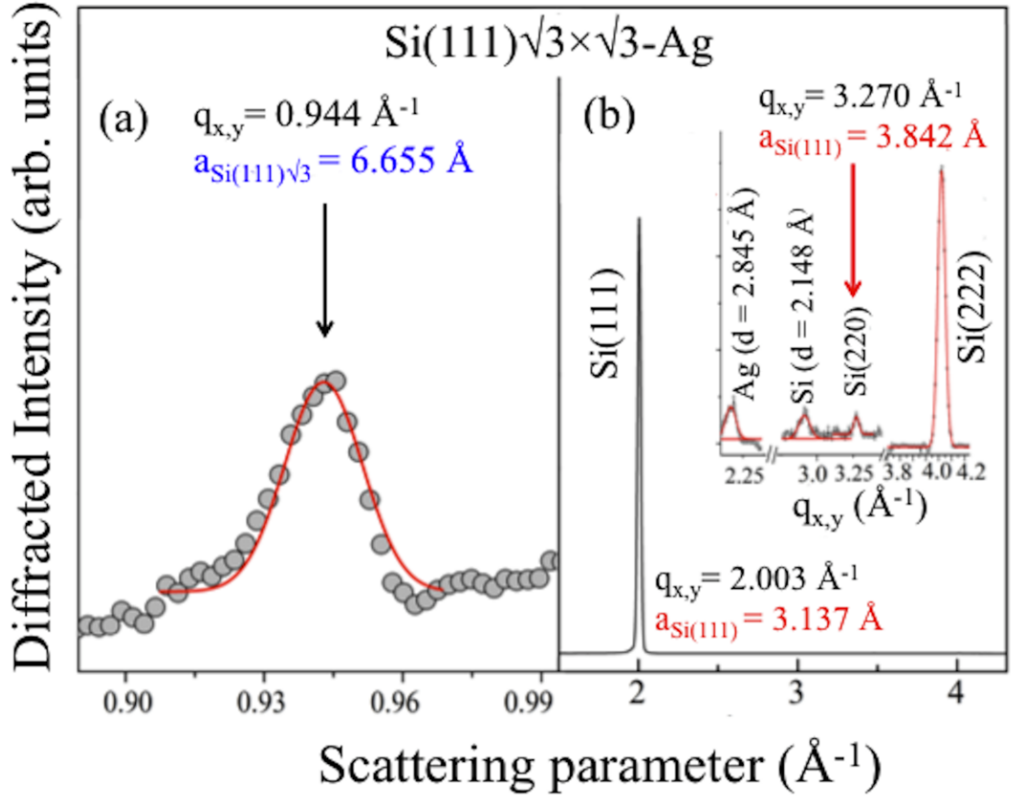

**Figure 2**. Energy Dispersive GIXRD pattern of the Si(111)√3×√3-Ag reference sample. (a) in-plane measurements. (b) The $q_z$ signal is dominant; the in-plane scattering vectors collected are shown in the inset. The diffraction peaks and their Gaussian fits (red line) are labelled: peaks positions are the centroid of the Gaussians and the full width at half maximum are: $FWHM_{Si(111)\sqrt{3}} = 0.0422$ Å$^{-1}$; $FWHM_{Ag} = 0.1868$ Å$^{-1}$, $FWHM_{Si-} = 0.130$ Å$^{-1}$; $FWHM_{Si(220)} = 0.086$ Å$^{-1}$; $FWHM_{Si(222)} = 0.065$ Å$^{-1}$.

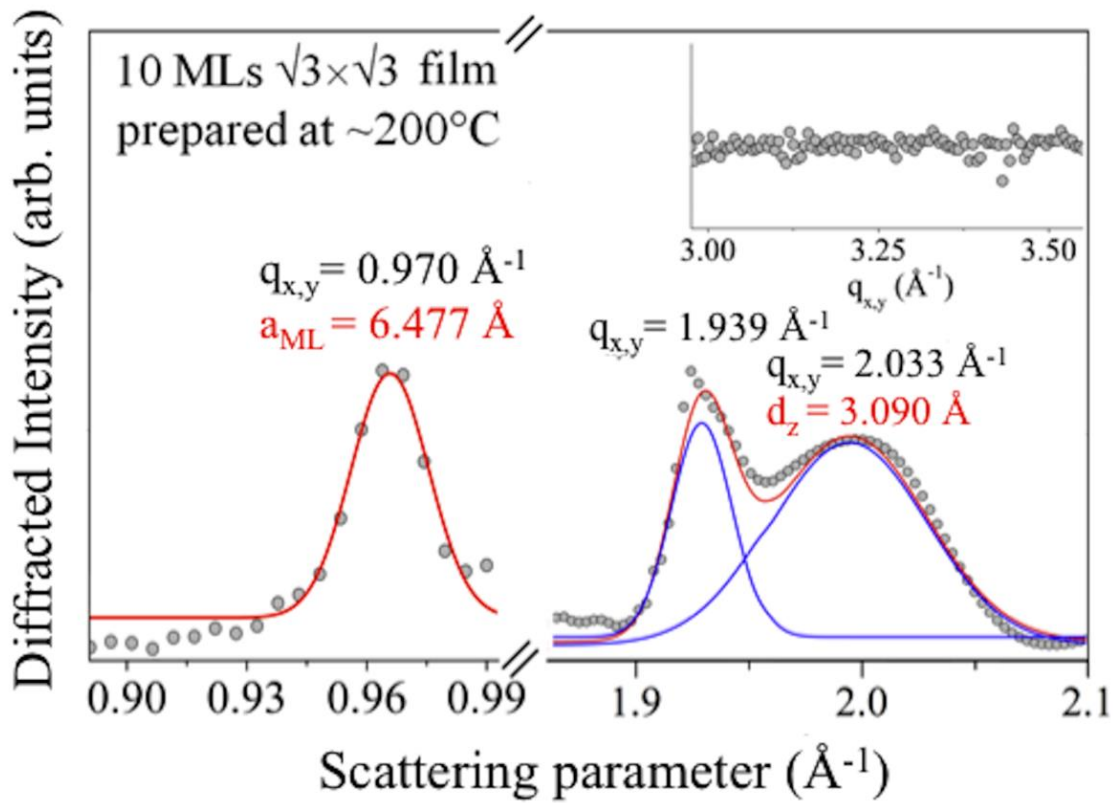

**Figure 3**. GIXRD pattern collected on a √3×√3 multilayer film (10 MLs) grown on Ag(111) at ~200°C and Gaussian fit (red line) of each reflection. The positions of the peaks are the centroid of the Gaussians; the full width at half maximum of each peak is reported. First order in-plane: $q_{xy}$ = 0.970 Å$^{-1}$ (FWHM$_{xy}$ = 0.0467 Å$^{-1}$) and out of plane $q_z$ reflections: $q_z$ = 2.033 Å$^{-1}$ (FWHM$_z$ = 0.2090 Å$^{-1}$); the second order in-plane reflection $q_{xy}$ = 1.939 Å$^{-1}$ (FWHM$_{xy}$ = 0.0790 Å$^{-1}$) is also detected (blue lines). The inset displays the GIXRD pattern around 3.27 Å$^{-1}$, as in Fig. 2: neither a Si(220) peak nor its relaxation are measured.

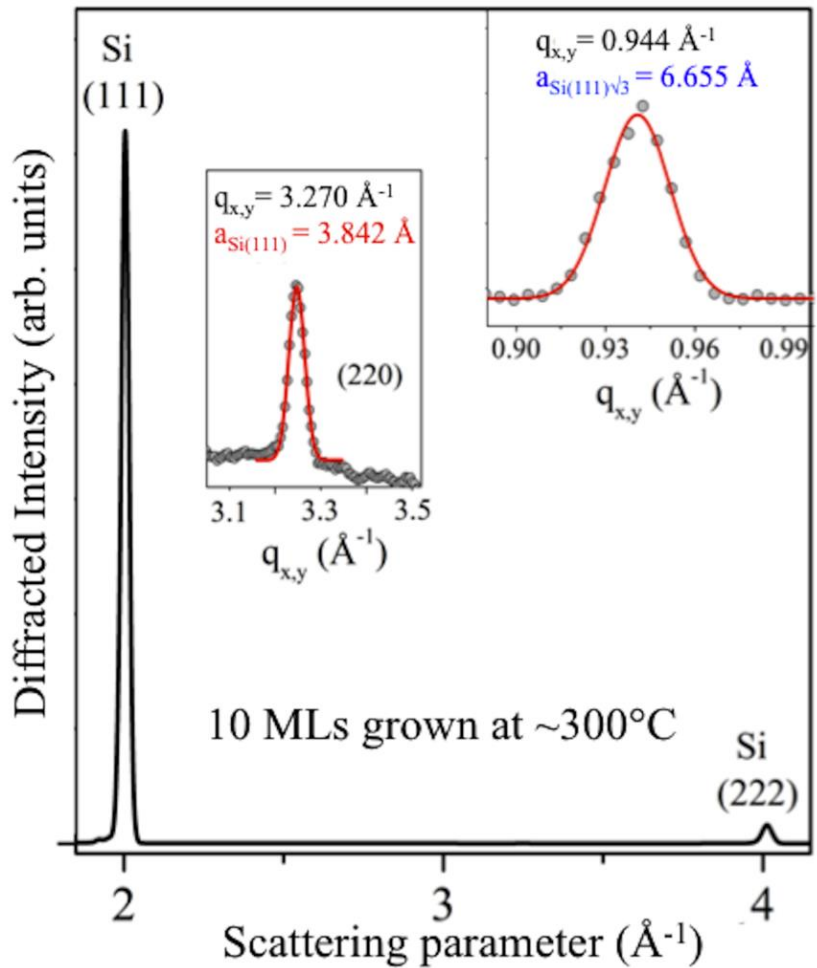

**Figure 4**. GIXRD pattern collected on a √3×√3 10 ML film grown on Ag(111) at ~300°C and Gaussian fit (red line) of each in-plane reflection. The positions of the peaks are the centroid of the Gaussians with full width half maximum of $FWHM_{Si(111)} = 0.0812 Å^{-1}$; $FWHM_{Si(111)\sqrt{3}} = 0.0520 Å^{-1}$.

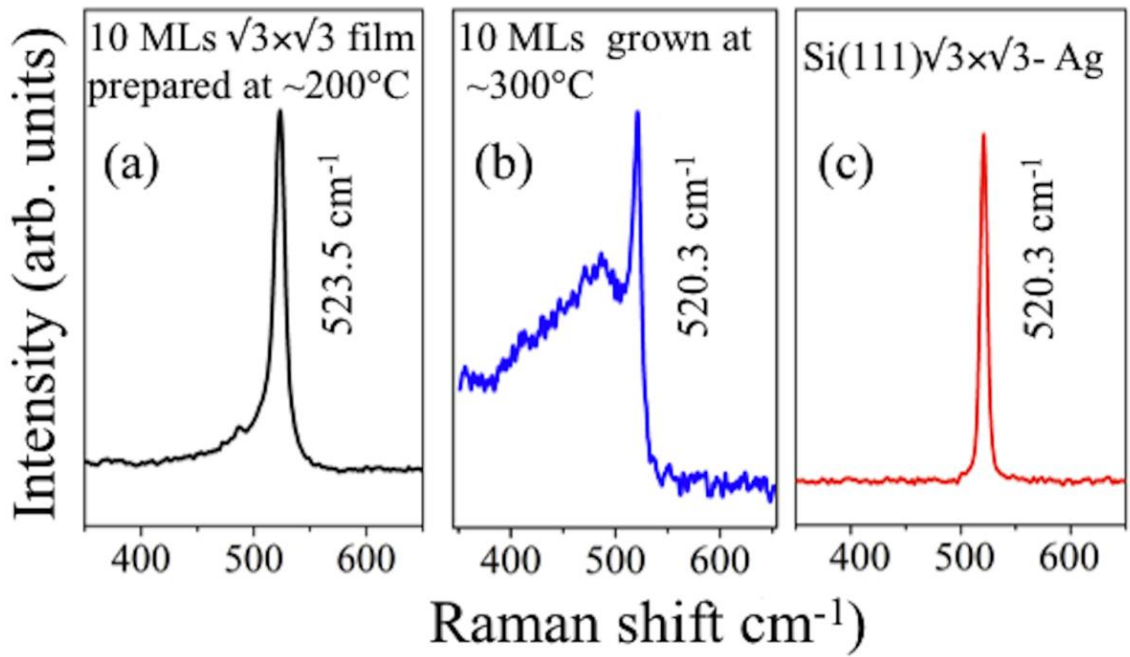

**Figure 5**. Ex-situ Raman spectroscopy measurements from (a) the 10 ML film grown on Ag(111) at ~200°C, (b) the 10 ML sample grown at ~300°C, showing the characteristic bulk silicon line at 520.3 cm$^{-1}$, and, (c) the reference Si(111)√3×√3-Ag sample.